\documentclass[aps,twocolumn,prb,superscript,floatfix,superscriptaddress,showpacs]{revtex4-2}
\usepackage{dcolumn}
\usepackage{bm}
\usepackage[skins,theorems]{tcolorbox}
\tcbset{highlight math style={enhanced,
  colframe=red,colback=white,arc=0pt,boxrule=1pt}}
\usepackage{color, graphicx}
\usepackage{braket,appendix,ushort}
\usepackage{accents}
\graphicspath{ {Images/} }

\begin{document}

\title{An improved auxiliary boson method for quantum dots with finite Coulomb interaction $U$} 

\author{J. P. Ramos-Andrade}
\email{juan.ramos@uantof.cl}
\affiliation{Departamento de F\'{\i}sica, Universidad de Antofagasta, Antofagasta, Chile.}
\author{P. A. Almeida}
\email{patriciaassisalmeida@usp.br}\affiliation{Instituto de Física, Universidade de São Paulo, C.P. 66318, 05315--970 São Paulo, SP, Brazil}
\author{A. M. Calle}
\email{acalle2@santotomas.cl}
\affiliation{Escuela de kinesiología, Facultad de Salud, Universidad Santo Tomás sede Antofagasta, Chile.}
\author{G. A. Lara}
\email{gustavo.lara@uantof.cl}
\affiliation{Departamento de F\'{\i}sica, Universidad de Antofagasta, Antofagasta, Chile.}

\date{\today}

\begin{abstract}
In this work, we present an improved version of the formulation of the four auxiliary bosons originally proposed by Kotliar and Ruckenstein, adapted to describe a quantum dot in the Kondo regime with finite electron–electron interaction $U$. The proposed refinement provides a more accu\-rate treatment of strong electronic correlations while preserving the analytical tractability of the original approach, thereby enabling a reliable description of transport properties in nanoscale systems. Benchmark calculations show good agreement with the results of the numerical renormalization group, supporting the validity of the improved formulation. Our approach therefore provides an efficient and robust framework for investigating transport through strongly correlated quantum dots.
\end{abstract}

\maketitle

\section{\label{introduction}Introduction}

In the interest of describing strong correlation phenomena that can arise in quantum dots (QDs), one analytical tool among others is the auxiliary boson formulation. The beauty of this tool lies in its simplicity; however, there are several different formulations, none of which can be claimed to be more correct than the others. The only guide is the agreement, in certain regions of the parameter space, with results from experiments, other theories, or numerical methods.

The first formulation of auxiliary particles was given by Barnes \cite{Barnes1976} for a magnetic impurity in an Anderson model, where the Coulomb term cannot be treated as a perturbation. To use Wick's theorem in its usual form \cite{FetterWalecka1971}, he introduced four auxiliary operators to represent the various singly occupied, doubly occupied, and unoccupied configurations in the impurity. Two of these operators are fermionic, $a_{\uparrow}$ and $a_{\downarrow}$, corresponding to the two singly occupied states, and the other two are bosonic, $b_{\uparrow \downarrow}$ and $b_0$, corresponding to the doubly occupied and unoccupied states, respectively. The reason why these last two operators are bosonic is to maintain a quadratic Hamiltonian form in fermionic operators. Finally, with this introduction of auxiliary operators, one can use temperature-ordered diagrams for the Schrieffer-Wolff limit of the Anderson model \cite{Barnes1976}.

Coleman \cite{Coleman1984} subsequently applied the idea of auxiliary particles to a formulation of the mixed-valence problem, in which the singlet valence state of a rare-earth ion is represented by a zero-energy boson and the spin state by a spin-$j$ fermion. This can be described by the infinite-$U$ Anderson model, where the double-occupancy state is forbidden, and thus only one auxiliary boson is needed. Additionally, the consideration of an auxiliary boson (which these authors call a ``slave boson") for this infinite-$U$ model was carried out by Read and Newns \cite{Read1983,Read1985}.
More recently, a unification between slave boson representation and the Keldysh path integral formalism has been considered to describe the Kondo effect in QDs \cite{PhysRevB.84.125303}. 
Earlier, Kotliar and Ruckenstein \cite{Kotliar1986} proposed the use of four auxiliary bosons to represent each of the electronic configurations of the sites in a Hubbard model.
An application of the auxiliary bosons proposed by Kotliar and Ruckenstein to linear and nonlinear transport through a QD with finite $U$ was carried out by Bing-Dong and Lei in Refs.~\cite{BingDong2001,BingDong2001b}, and in a double QD system~\cite{BingDong2002}. In addition, this method was also applied to a mesoscopic ring coupled with a QD~\cite{Ding2003}, a QD molecule~\cite{VERNEK2006608}, and in a QD connected to ferromagnetic leads~\cite{JingMa_2004}.

In this work, we aim to present an improved version of the four auxiliary boson method proposed by Kotliar and Ruckenstein, specifically adapted for QDs that predominantly operate in the Kondo regime. By refining this approach, we seek to enhance the accuracy and applicability of the method in describing strong correlation effects in nanoscale systems. This development could provide valuable insights into transport properties and correlation phenomena in such systems, with potential implications for both theoretical modeling and experimental studies.

This paper is organized as follows: Section\ \ref{secII} introduces the model Hamiltonian; Section\ \ref{secIII} presents the treatment using the auxiliary bosons the corresponding results and discussion; Section~\ref{secIV} derives the Green’s functions used to calculate the relevant electronic mean values. Finally, Section~\ref{secV} provides the concluding remarks.

\section{Hamiltonian model}\label{secII}

\begin{figure}[h]
 \centering
 \includegraphics[width=.3\textwidth]{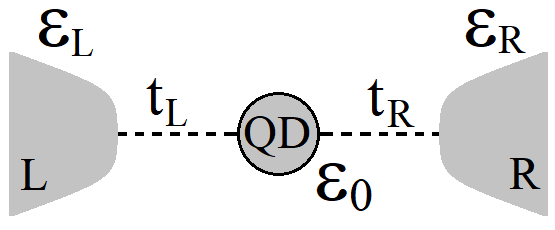}
  \caption{Schematic view of the setup: the QD coupled to left/right leads. The parameter $t_{\text{L(R)}}$ (dashed lines) denotes the leads-QD coupling. }
  \label{fig1}
\end{figure}

The system under study is the one presented in FIG.~\ref{fig1}: a single-level QD connected to two reservoirs, one on the left (L), acting as a source, and one on the right (R), acting as a drain for electric current transport.
The system is model by a low-energy Hamiltonian $H$ in the form
\begin{equation}
 H = H_{\text{leads}} +H_{\text{QD}} +H_{\text{I}}\,,
\end{equation}
where the terms: $H_\text{leads}$ corresponds to normal leads (L and R);
$H_{\text{QD}}$ corresponds to the QD; and $H_{\text{I}}$ to the coupling 
between leads and the QD.
These are, respectively, given by
\begin{equation} 
H_\text{leads} =  
\sum_{\substack{\alpha = \{ \text{L},\text{R}\} \\ k_{\alpha} \in \alpha }}
\sum_{\sigma = \{ \uparrow , \downarrow \}}    \varepsilon_{k_{\alpha}} 
     c_{k_{\alpha}\sigma}^{\dag}  c_{k_{\alpha}\sigma}\,, 
\end{equation} 

\begin{equation} 
H_{\text{QD}} =  
\sum_{\sigma = \{ \uparrow , \downarrow \}} \varepsilon_{0\sigma} \hat{n}_{0\sigma} 
+U \hat{n}_{0\uparrow} \hat{n}_{0\downarrow}\,,
\end{equation} 

\begin{equation} 
H_{I} =  
-\! \! \sum_{\substack{\alpha = \{ \text{L},\text{R}\} \\ k_{\alpha} \in \alpha  }}
 \sum_{\sigma=\{ \uparrow , \downarrow \} }               
  t_\alpha \left[ c_{k_\alpha \sigma}^\dag c_{0\sigma}
                + c_{0\sigma}^{\dag} c_{k_\alpha \sigma}  \right]\,,
\end{equation} 
where $c_{k_{\alpha}\sigma}^{\dag}$ $(c_{k_{\alpha}\sigma})$ creates (annihilates) an electron with momentum $k_{\alpha}$, spin $\sigma=\,\uparrow$ or $\downarrow$, and energy $\varepsilon_{k_{\alpha}}$ in the lead $\alpha=\text{L},\text{R}$. Also, $c_{0\sigma}^{\dag}$ $(c_{0\sigma})$ creates (annihilates) an electron with energy $\varepsilon_{0\sigma}$ in the single energy level of the QD corresponding to the spin $\sigma$, $\hat{n}_{0\sigma}$ is the corresponding occupation-number operator, and $U$ represents the on-site Coulomb repulsion energy associated with double occupancy of the QD by electrons with opposite spins.

\section{Treatment of the Coulomb interaction: auxiliary bosons}\label{secIII}

To treat the Coulomb interactions, auxiliary bosons 
are used, similarly to the approach developed by Kotliar and Ruckenstein 
\cite{Kotliar1986}.
The system is extended to include, in addition to the electronic degrees of freedom, 
four types of bosons associated with each of the four possible 
fermionic states of the QD. Specifically, $\hat{e}^{\dag}$ ($\hat{e}$) is the 
bosonic operator associated with the creation (annihilation) of the empty state in the QD; 
$\hat{s}_{\sigma }^{\dag}$ ($\hat{s}_{\sigma}$) is the bosonic operator 
associated with the creation (annihilation) of a singly occupied state with spin $\sigma$ in the QD; and $\hat{d}^{\dag}$ ($\hat{d}$) is the bosonic operator 
associated with the creation (annihilation) of the doubly occupied state in the QD.

\noindent 
In the Hamiltonian of this extended system, the bosons are used to represent the Coulomb interaction, and the dynamics of the electrons is coupled to the dynamics of the bosons. A constraint imposed on the auxiliary bosons is that the number of each type of boson must correspond to the probability of occupation of the corresponding electronic state. Thus, the sum over the four bosonic occupation numbers must equal one (completeness condition), as summarized by:
\begin{equation}\label{restriction_completitude}
 |\hat{e}|^2 + |\hat{s}_{\uparrow }|^2  
 + |\hat{s}_{\downarrow}|^2 +|\hat{d}|^2  - 1 = 0\,.
\end{equation}
Two other restrictions must be considered: the number of electrons in the electronic space should be equivalent to the number in the auxiliary bosonic subspaces, then
\begin{equation}  \label{restriction_ni}
 c_{0\sigma}^{\dag} c_{0\sigma} - |\hat{s}_{\sigma }|^2 -|\hat{d}|^2
= 0\,.
\end{equation}
 \noindent 
 And the transfer of electrons to or from the QD in the electronic subspace must also be equivalent in the bosonic subspace, thus
\begin{equation} \label{restriction_x0}
\sum_{\alpha,k_{\alpha}} t_{\alpha}
\left( c^{\dag}_{k_{\alpha} \sigma}
 c_{0\sigma}
-
\sqrt{\frac{\langle \hat{n}_{k_{\alpha} \sigma} \rangle }{N}}
\left[ \hat{s}_{ \bar{\sigma}}^{\dag} \hat{d} +
        \hat{e}^{\dag} \hat{s}_{\sigma}
  \right]
\right)
  = 0\,,
\end{equation}
where, here and throughout the rest of the document, $\bar{\sigma}=-\sigma$. This operator on the bosonic spaces corresponds to two possible transitions associated with the destruction of the electron. 
Then, using the constraints given by Eqs.~(\ref{restriction_completitude}),~(\ref{restriction_ni}),  and~(\ref{restriction_x0}), with the Lagrange multipliers $\lambda_{0}$, $\lambda_{n\sigma}$,  and $\lambda_{t\sigma}$, respectively, we obtain an extended Hamiltonian given by

\begin{equation}
\begin{aligned} 
H_{\text{ext}} = &
\sum_{\alpha,\sigma, k} 
\Biggl( {\varepsilon}_{k} c_{k\sigma}^\dag c_{k\sigma} 
      -t_\alpha (1-\lambda_{t\sigma}) \left( c_{k\sigma}^\dag  c_{0\sigma}
          +\text{h.c.} \right) 
\\ &  
+\left( \varepsilon_{0} +\lambda_{n\sigma} \right) c_{0\sigma}^{\dag} c_{0\sigma} 
 -\lambda_{n\sigma} 
                \left(  |\hat{s}_{\sigma}|^2 +|\hat{d}|^2 \right) 
\\ &
-\lambda_{t\sigma} 
 t_{\alpha}
\sqrt{\frac{\langle \hat{n}_{k_{\alpha} \sigma} \rangle }{N}}
\left[ \hat{s}_{ \bar{\sigma}}^{\dag} \hat{d} +
        \hat{e}^{\dag} \hat{s}_{\sigma}
 +\text{h.c.}  \right]\Biggr) \\ &
 + U |\hat{d}|^2 
 +\lambda_{0} \left( |\hat{e}|^2 |+|\hat{s}_{\uparrow}|^2 
   + |\hat{s}_{\downarrow}|^2 +|\hat{d}|^2-1 \right)
\,.
\end{aligned} 
\end{equation}

Now we consider that the temperature is low enough such that fluctuations of the bosonic operators around their mean values are negligible. The mean values of the bosonic number operators can then be replaced by real numbers, thus

\begin{equation}
 b = \langle \hat{b}^{\dag} \rangle 
        = \langle \hat{b} \rangle 
        \,, \; b=\left\{ e,\, s_{\uparrow},\, s_{\downarrow}, \,d \right\}\,.
\end{equation}
Finally, we have the following one-particle mean-field Hamiltonian
\begin{equation} \label{Hmf}
\begin{aligned} 
 H_{\text{mf}}= & \sum_{\sigma} 
 \left\{ \begin{aligned} 
 & \sum_{\alpha,k} \left[
 \varepsilon_{k} \hat{n}_{k\sigma}
 -\tilde{t}_{\alpha} \left( c_{k\sigma}^{\dag} c_{0\sigma} +\text{h.c.} \right)
 \right]
 \\ & 
 +\widetilde{\varepsilon}_{0} \hat{n}_{0\sigma}
 \end{aligned} \right\} 
 \\ &
+ U d^2 
+\lambda_0 \left( e^2+s_{\uparrow}^2 +s_{\downarrow}^2 +d^2 -1\right)
\\ & \hspace{-1em} 
-\sum_{\sigma} 
\left[ \lambda_{n\sigma} \left(d^2 +s_{\sigma}^2 \right)
+\lambda_{t\sigma} 
h_{\sigma} (2es_{\sigma} +2ds_{\bar{\sigma}}) \right] , \end{aligned} 
\end{equation}
where
$\tilde{t}_{\alpha} = 
t_{\alpha} (1-\lambda_{t\sigma} )$,
$\widetilde{\varepsilon}_{0} =  \varepsilon_{0} +\lambda_{n\sigma}$, and $h_{\sigma}=\sum_{\ell} t_{\ell}\sqrt{\hat{n}_{\ell}/N}$.

Both $\tilde{t}_{\alpha}$ and $\widetilde{\varepsilon}_{0} $ are determined by two conditions that we have obtained from the analysis of NRG results and which we show in the next section. In this way, the following average electronic values are also determined, 
\begin{equation} 
\bar{n}_{0\sigma} 
=  \left\langle \hat{n}_{0\sigma} \right\rangle 
=  \left\langle c_{0,\sigma}^{\dag} c_{0,\sigma} \right\rangle\,.
\end{equation}

\begin{equation}
\bar{\xi}_{0\sigma}
  =
  \sum_{\alpha,k_\alpha} \tilde{t}_{\alpha \sigma}
 \left\langle c_{k_\alpha \sigma}^{\dag} c_{0\sigma}
        +c_{0\sigma}^{\dag} c_{k_\alpha\sigma}
 \right\rangle.
\end{equation} 

\noindent 
The four bosonic occupancy parameters, $e$, $d$, $s_{\uparrow}$, $s_{\downarrow}$, can be determined by minimizing the ground state energy of the mean-field Hamiltonian. The conditions for minimal energy, together with the application of the Hellmann–Feynman theorem ($\partial\langle H\rangle/\partial x  = \langle \partial H/\partial x \rangle$) to the Hamiltonian, give the following set of self-consistent equations

\begin{equation} \label{completitude}
\left\langle
\frac{\partial H_{\text{eff}}}{\partial \lambda_{0}}
\right\rangle
 = 
 e^{2} +s_{\uparrow}^2 +s_{\downarrow}^2 +d^2 - 1  = 0\,, 
\end{equation}

\begin{equation} \label{numero}
\left\langle
\frac{\partial H_{\text{eff}}}{\partial \lambda_{n\sigma}}
\right\rangle
 = 
 \langle c^{\dag}_{0\sigma} c_{0\sigma} \rangle 
 -\left(  s_{\sigma}^{2} +d^{2} \right)    = 0 \,,
\end{equation}

\begin{equation} \label{transfer}
\left\langle
\frac{\partial H_{\text{eff}}}{\partial \lambda_{t\sigma}}
\right\rangle
 = 
 2\sum_{\ell} \widetilde{t}_{\ell}
 \left\langle c^{\dag}_{\ell \sigma} c_{0\sigma} \right\rangle 
 -2 h_{\sigma} 
 \left( e s_{\sigma} +s_{\bar{\sigma}} d \right)
 = 0 \,,
\end{equation}
      
This gives us the system of equations

\begin{align}
 e^2+s_{\sigma}^2+s_{\bar{\sigma}}^2+d^2 = & 1
\\  
 s_{\sigma}^{2} +d^{2} = & \bar{n}_{0\sigma}
\\  
2 s_{\sigma} e +2 s_{\bar{\sigma}} d
= &
q_{\sigma}
\end{align}
where
\begin{equation}
 q_{\sigma} = \frac{\bar{\xi}_{0\sigma}}{(1-\lambda_{t\sigma}) h_{\sigma}}.
\end{equation}

\subsubsection{Spin symmetry case}
Spin symmetry imposes the condition 
$s_{\uparrow}=s_{\downarrow} = s$, and the solutions for occupancy probabilities are
\begin{equation} 
s^2 = \frac{q_{\sigma}^2}{4}
\frac{1
+\sqrt{ 4(1-\bar{n}_{0\sigma}) \bar{n}_{0\sigma} -q_{\sigma}^2 } }
{q_{\sigma}^2 +\left[1-4(1-\bar{n}_{0\sigma})  \bar{n}_{0\sigma}\right] }\,,
\end{equation} 

\begin{equation} 
e^2 = 1-\bar{n}_{0\sigma}
-\frac{q_{\sigma}^2}{4}
\frac{1
+\sqrt{ 4(1-\bar{n}_{0\sigma}) \bar{n}_{0\sigma} -q_{\sigma}^2 } }
{q_{\sigma}^2 +\left[1-4(1-\bar{n}_{0\sigma})  \bar{n}_{0\sigma}\right] }\,,
\end{equation} 

\begin{equation} 
d^2 = \bar{n}_{0\sigma}
-\frac{q_{\sigma}^2}{4}
\frac{1
+\sqrt{ 4(1-\bar{n}_{0\sigma}) \bar{n}_{0\sigma} -q_{\sigma}^2 } }
{q_{\sigma}^2 +\left[1-4(1-\bar{n}_{0\sigma})  \bar{n}_{0\sigma}\right] }\,.
\end{equation} 

\subsubsection{Electron-hole symmetry case}
In addition to spin symmetry, $s_{\uparrow}=s_{\downarrow} = s$, in this case there is electron-hole symmetry, that is,
$\varepsilon_{0} = -U/2$. From the analysis of the NRG results, the condition for the renormalized energy in the QD is $\widetilde{\varepsilon}_0 = 0$, which in turn determines that $\bar{n}_{0\sigma} = 1/2$.

The solutions are
\begin{equation}
 s^2 = \frac{1}{4}
 \left[ 1 +\sqrt{1-q_{\sigma}^2} \right]\,,
\end{equation}

\begin{equation}
 e^2 = d^2 = \frac{1}{4}
 \left[ 1 -\sqrt{1-q_{\sigma}^2} \right]\,.
\end{equation}

\begin{figure}[htb!]
\centering
\includegraphics[width=0.5\textwidth]{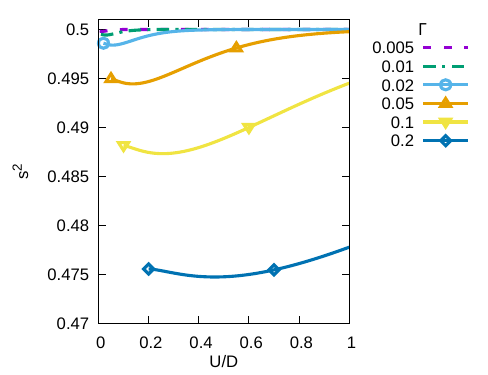}
\caption{ Probability that the QD is in singly occupied state in the case of electron-hole symmetry.
\label{s2_vs_U_Symmetric}
}
\end{figure}

\begin{figure}[htb!]
\centering
\includegraphics[width=0.5\textwidth]{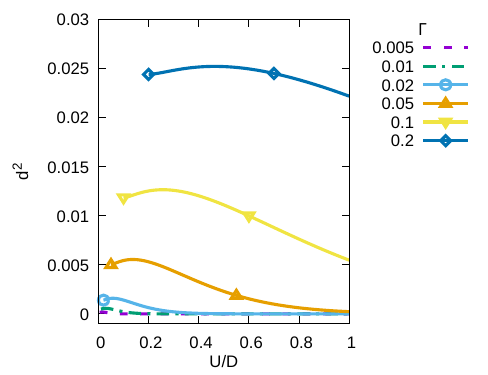}
\caption{ Probability that the QD is in the doubly occupied state or in the empty state, in the case of electron-hole symmetry.
\label{d2_vs_U_Symmetric}
}
\end{figure}

\subsection{Kondo temperature and NRG} 
A QD connected to metallic leads is commonly described by the non-degenerate Anderson impurity model. When the single energy level of the QD, $\varepsilon_0$, is adjusted away from mixed valence regimes, the Kondo effect manifests. This phenomenon involves the interaction between the localized unpaired spins of the QD and the conduction electrons of the metallic leads, culminating in the formation of a non-magnetic state at sufficiently low temperatures. A key signature of this state is the emergence of an Abrikosov-Suhl resonance below a characteristic Kondo temperature, $T_K$ \cite{Tsvelick1983,Pruschke1989}, whose behavior is given by

\begin{equation} \label{T_K}
 \kappa_B T_K = \frac{\text{min}(U,D)}{2\pi} \sqrt{I} e^{-\pi/I},
\end{equation}
where $\kappa_{B}$ is the Boltzmann constant,
\begin{equation}\label{II}
I = 2\Gamma
 \left[ \frac{1}{|\varepsilon_{0}|} +\frac{1}{U+\varepsilon_{0}} \right] \,,
\end{equation}

\noindent
and $\Gamma = \pi \sum_{\alpha}|t_{\alpha}|^2 \rho_{\alpha}$ denotes the hybridization between the localized level and the conduction electrons in the leads. This quantity is characterized as the HWHM of the resonance at the Fermi level, with the leads having a density of states $\rho_{\alpha}$. The expression for the Kondo temperature given by Eq.~(\ref{T_K}) is valid provided that the condition $\Gamma \ll {|\varepsilon_0|, U}$ is satisfied.

\begin{figure}[htb!]
\centering
\includegraphics[width=0.5\textwidth]{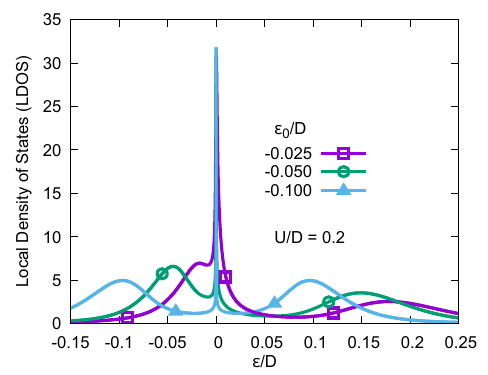}
\caption{\label{LDOS_NRG_1} LDOS as a function of energy for several values of $\varepsilon_0$, obtained from NRG. Here, $D$ is the energy unit ($D = 1$),  $U = 0.2D$, $\Gamma = 0.01D$, and the temperature is such that $\kappa_B T = 10^{-15}D$ .
\label{fignrg}
}
\end{figure}

To validate the results of the improved auxiliary boson approach presented, the NRG technique is employed for comparison. A non-perturbative method for systems with magnetic impurities, NRG begins with the logarithmic discretization of the electron bath, mapped onto a 1D tight-binding chain, and proceeds iteratively by incorporating new sites into the impurity–chain system. At each step, the Hamiltonian is diagonalized, retaining only low-energy states while discarding high-energy ones according to the scale separation principle. The resulting flow of rescaled energy levels between fixed points highlights the renormalization character of the method, which efficiently captures the relevant energy scales from the bandwidth, $D$, down to the $T_K$.
 Our calculations are conducted with  the Ljubljana open-source code  \cite{zitko_rok}. A discretization parameter of $\Lambda = 2.0$ was applied, ensuring sufficient resolution and maintaining a minimum of 10.000 states at each iterative step. Furthermore, the implementation of the $z$-trick \cite{Campo2005}, with $z$ values ranging from 0.0625 to 1.0 (i.e., $N_z = 4$), demonstrated a stabilizing effect, thereby mitigating oscillations and artifacts in the computed physical quantities. The impurity local density of states (LDOS) was derived using the density matrix numerical renormalization group (DM-NRG) approximation \cite{Hofstetter2000}.

In the NRG framework, the QD LDOS exhibits three characteristic peaks, as shown in FIG.~\ref{LDOS_NRG_1}. The central peak (located near the Fermi energy) corresponds to spin fluctuations and can be compared with the LDOS obtained from the auxiliary boson method presented above.

\begin{figure}[htb!]
\centering
\includegraphics[width=0.5\textwidth]{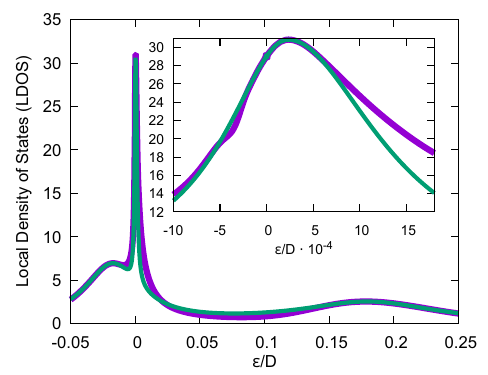}
\caption{\label{LDOS_Fit}
LDOS as a function of energy. The NRG results (thick purple line) are fitted to three asymmetric Lorentzian curves (thin green line) to obtain a more accurate fit of the side of the central peak closest to the Fermi level ($\varepsilon =0$). These curves reproduce the NRG results at the energy corresponding to each peak maximum and in adjacent energy regions. The inset shows a detailed comparison of the central peak. }\label{fig5}
\label{fignrg}

\end{figure}

Using NRG, the local density of states of the QD is obtained for various values of $U$, $\varepsilon_0$, and $\Gamma$. These results are fitted by three Lorentzian curves where the central peak is considered asymmetric, as shown in FIG.~\ref{fig5}. In this way, we obtain empirical results for a Lorentzian that best fits the values around the Fermi level and which we will identify as the Abrikosov-Suhl resonance below $T_K$.

\noindent
Thus, we have empirically obtained two quantities: the FWHM, $\Delta$, and the energy at which the maximum occurs, $\widetilde{\varepsilon}_0$. 
The FWHM has the following simple expression, 
\begin{equation} \label{DeltaNRG}
\begin{aligned} 
\Delta
= &
\frac{U}{2} 
\sqrt{ I \frac{ 1-\left(\frac{U+2\varepsilon_0}{U-2\Gamma} \right)^2 }{1+\frac{U}{D} } }
\exp{\left\{ \frac{\Gamma}{U} \left[ 1+\frac{U}{D}\right] -\frac{\pi}{I} \right\} }. 
\end{aligned} 
\end{equation}

\noindent
On the other hand, the energy corresponding to the maximum of the resonance peak has the following  expression

\begin{equation} \label{E0eff}
\begin{aligned} 
\widetilde{\varepsilon}_0
= &
\frac{ (U+2\varepsilon_0) }{\left( U+\frac{\Gamma}{4} \right)^{3/2} } 
\; \frac{ I\, U \, \sqrt{I\, \Gamma} }{2\pi}  
\\ &
\exp{ \left\{ 0.6 \left[ 1+\frac{U}{D} \right] -\frac{9\Gamma}{D} -\left| \frac{U+2\varepsilon_0}{D} \right| -\frac{\pi}{I} \right\} }\,.
\end{aligned} 
\end{equation}

\noindent
In both expressions, $I$ correspond to the definitions given in Eq.~(\ref{II}).

Finally, within the regime restrictions $U<2D$, and $U,|\tilde{\varepsilon}_0|>2\Gamma$, comparison between the results obtained using the proposed method and the NRG approach shows that the relative deviation does not exceed $2.0\%$ for $\Delta$ and $3.5\%$ for $\tilde{\varepsilon}_0$. These results demonstrate the good quantitative accuracy of the method within its range of applicability and support its use as a practical tool for further investigations of strongly correlated QD systems.

\section{Green's functions and mean values }\label{secIV}

The electronic mean values are obtained using the Green's function method applied to the mean-field single-particle Hamiltonian shown in Eq.~(\ref{Hmf}). 

These quantities can be expressed in terms of the lesser Green's functions as follows:

\begin{equation}
\bar{n}_{0\sigma } 
 = 
 \int\limits_{-\infty}^{\infty} \frac{\text{d}\varepsilon}{2\pi} \left[ -i 
G_{00}^{\sigma,<} (\varepsilon) \right],
\end{equation}
and
\begin{equation}
\bar{\xi}_{0 \sigma}
 =
  \int\limits_{-\infty}^{\infty} \frac{d\varepsilon}{2\pi}
 \sum_{\alpha,k_{\alpha}} \tilde{t}_{\alpha}
  \left[ -i G_{k_{\alpha} 0}^{\sigma,<} (\varepsilon) \right]\,.
\end{equation}

Using the density of states of a semi-infinite chain,
\begin{equation}
 {\cal N}(\varepsilon_k) = \frac{2}{\pi D} \sqrt{1 -(\varepsilon_{k}/D)^2}\,,
\end{equation}
these Green's functions have non-zero values within the energy band $-D \leq \varepsilon \leq D$ and it can be shown that the Green's functions reduce to the expressions

\begin{equation}
G_{00}^{\sigma,<} =  
 \frac{i 2 \Sigma_I {\frak f}(\varepsilon ) }
 {\left[\varepsilon -\tilde{\varepsilon}_{0} -\Sigma_R \right]^2 
 +\Sigma_{I}^2 },
\end{equation}

\begin{equation}
\sum_{\alpha,k} \tilde{t}_{\alpha} G_{k0}^{\sigma,<} =  
 \frac{i 2 \Sigma_I {\frak f}(\varepsilon ) \, (\varepsilon -\tilde{\varepsilon}_{0} ) }
 {\left[\varepsilon -\tilde{\varepsilon}_{0}-\Sigma_R \right]^2 
 +\Sigma_{I}^2 },
\end{equation}
where $\Sigma_R =\varepsilon/D$, 
$\Sigma_I= \widetilde{\Gamma}_{\sigma} \sqrt{1-\left(\varepsilon/D\right)^2} $,  
$ {\frak f}(\varepsilon )$ is the Fermi's distribution function, and
\[
\widetilde{\Gamma} = \pi {\cal N}(0) 
\sum_{\ell} t_{\ell}^2 
\]
is the half-width at half height of the effective resonance.

Then, at zero temperature
\begin{equation}
 \bar{n}_{0\sigma }
 =
 \frac{1}{\pi} \gamma \int\limits_{0}^{1}
 \frac{  dx \, \sqrt{1-x^2} }
  {\left(x -a -\gamma x \right)^2 +\gamma^2 (1-x^2)  }\,,
\end{equation}

\begin{equation}
  \xi_{0 \sigma}
   =
 -D \frac{1}{\pi}  \gamma
 \int\limits_{0}^{1}
  \frac{ dx \, \sqrt{1-x^2} (x-a ) }
       { \left( x-a -\gamma x \right)^2 +\gamma^2 (1-x^2)    }\,,
\end{equation}
where we have defined $\gamma = \widetilde{\Gamma}/D$, and $a=\widetilde{\varepsilon}_0/D $.

\section{Final Remarks}\label{secV}

In this work, we have presented a modification of the Kotliar–Ruckenstein auxiliary boson method, adapted to describe the electronic properties of QDs in the finite-$U$ Kondo regime. By reformulating the mean-field treatment, analytical expressions for Green functions, and associated quantities of interest, are obtained. We derived an effective Hamiltonian to characterize key features of strong electron correlations in QD-leads systems. We obtained the lesser Green’s functions, which allowed us to compute the average electronic values. The resulting spectral function exhibits a Lorentzian profile whose width agrees with the expected Kondo temperature dependence. This provides strong evidence that the improved method is consistent with benchmark results from numerical approaches such as the NRG. Our adapted treatment retains the analyticity of the auxiliary boson approach and enhances its accuracy in regimes where many-body correlations play a dominant role. Then, the presented results can be understood as a tool for modeling transport properties in strongly correlated nanoscale systems and could be extended to nonequilibrium scenarios or multiple QDs configurations.

\acknowledgments

J.P.R.-A. is grateful for the financial support of FONDECYT Iniciaci\'on grant No. 11240637. J.P.R.-A. and G.A.L. acknowledge the financial support of the MINEDUC-UA project, code ANT22991. P. A. Almeida thanks FAPESP (Grant No. 2025/21932-6).

\bibliographystyle{apsrev4-2}  
\bibliography{referencias}

\begin{thebibliography}{18}%
\makeatletter
\providecommand \@ifxundefined [1]{%
 \@ifx{#1\undefined}
}%
\providecommand \@ifnum [1]{%
 \ifnum #1\expandafter \@firstoftwo
 \else \expandafter \@secondoftwo
 \fi
}%
\providecommand \@ifx [1]{%
 \ifx #1\expandafter \@firstoftwo
 \else \expandafter \@secondoftwo
 \fi
}%
\providecommand \natexlab [1]{#1}%
\providecommand \enquote  [1]{``#1''}%
\providecommand \bibnamefont  [1]{#1}%
\providecommand \bibfnamefont [1]{#1}%
\providecommand \citenamefont [1]{#1}%
\providecommand \href@noop [0]{\@secondoftwo}%
\providecommand \href [0]{\begingroup \@sanitize@url \@href}%
\providecommand \@href[1]{\@@startlink{#1}\@@href}%
\providecommand \@@href[1]{\endgroup#1\@@endlink}%
\providecommand \@sanitize@url [0]{\catcode `\\12\catcode `\$12\catcode `\&12\catcode `\#12\catcode `\^12\catcode `\_12\catcode `\%12\relax}%
\providecommand \@@startlink[1]{}%
\providecommand \@@endlink[0]{}%
\providecommand \url  [0]{\begingroup\@sanitize@url \@url }%
\providecommand \@url [1]{\endgroup\@href {#1}{\urlprefix }}%
\providecommand \urlprefix  [0]{URL }%
\providecommand \Eprint [0]{\href }%
\providecommand \doibase [0]{https://doi.org/}%
\providecommand \selectlanguage [0]{\@gobble}%
\providecommand \bibinfo  [0]{\@secondoftwo}%
\providecommand \bibfield  [0]{\@secondoftwo}%
\providecommand \translation [1]{[#1]}%
\providecommand \BibitemOpen [0]{}%
\providecommand \bibitemStop [0]{}%
\providecommand \bibitemNoStop [0]{.\EOS\space}%
\providecommand \EOS [0]{\spacefactor3000\relax}%
\providecommand \BibitemShut  [1]{\csname bibitem#1\endcsname}%
\let\auto@bib@innerbib\@empty
\bibitem [{\citenamefont {Barnes}(1976)}]{Barnes1976}%
  \BibitemOpen
  \bibfield  {author} {\bibinfo {author} {\bibfnamefont {S.~E.}\ \bibnamefont {Barnes}},\ }\href@noop {} {\bibfield  {journal} {\bibinfo  {journal} {J. Phys. F: Metal Phys.}\ }\textbf {\bibinfo {volume} {6}},\ \bibinfo {pages} {1375} (\bibinfo {year} {1976})}\BibitemShut {NoStop}%
\bibitem [{\citenamefont {Fetter}\ and\ \citenamefont {Walecka}(1971)}]{FetterWalecka1971}%
  \BibitemOpen
  \bibfield  {author} {\bibinfo {author} {\bibfnamefont {A.~L.}\ \bibnamefont {Fetter}}\ and\ \bibinfo {author} {\bibfnamefont {J.~D.}\ \bibnamefont {Walecka}},\ }\href@noop {} {\emph {\bibinfo {title} {Quantum Theory of Many-Particle Systems}}}\ (\bibinfo  {publisher} {McGraw-Hill},\ \bibinfo {address} {New York},\ \bibinfo {year} {1971})\BibitemShut {NoStop}%
\bibitem [{\citenamefont {Coleman}(1984)}]{Coleman1984}%
  \BibitemOpen
  \bibfield  {author} {\bibinfo {author} {\bibfnamefont {P.}~\bibnamefont {Coleman}},\ }\href@noop {} {\bibfield  {journal} {\bibinfo  {journal} {Phys. Rev. B}\ }\textbf {\bibinfo {volume} {29}},\ \bibinfo {pages} {3035} (\bibinfo {year} {1984})}\BibitemShut {NoStop}%
\bibitem [{\citenamefont {Read}\ and\ \citenamefont {Newns}(1983)}]{Read1983}%
  \BibitemOpen
  \bibfield  {author} {\bibinfo {author} {\bibfnamefont {N.}~\bibnamefont {Read}}\ and\ \bibinfo {author} {\bibfnamefont {D.~M.}\ \bibnamefont {Newns}},\ }\href@noop {} {\bibfield  {journal} {\bibinfo  {journal} {J. Phys. C}\ }\textbf {\bibinfo {volume} {16}},\ \bibinfo {pages} {3273} (\bibinfo {year} {1983})}\BibitemShut {NoStop}%
\bibitem [{\citenamefont {Read}(1985)}]{Read1985}%
  \BibitemOpen
  \bibfield  {author} {\bibinfo {author} {\bibfnamefont {N.}~\bibnamefont {Read}},\ }\href@noop {} {\bibfield  {journal} {\bibinfo  {journal} {J. Phys. C}\ }\textbf {\bibinfo {volume} {18}},\ \bibinfo {pages} {2651} (\bibinfo {year} {1985})}\BibitemShut {NoStop}%
\bibitem [{\citenamefont {Smirnov}\ and\ \citenamefont {Grifoni}(2011)}]{PhysRevB.84.125303}%
  \BibitemOpen
  \bibfield  {author} {\bibinfo {author} {\bibfnamefont {S.}~\bibnamefont {Smirnov}}\ and\ \bibinfo {author} {\bibfnamefont {M.}~\bibnamefont {Grifoni}},\ }\href {https://doi.org/10.1103/PhysRevB.84.125303} {\bibfield  {journal} {\bibinfo  {journal} {Phys. Rev. B}\ }\textbf {\bibinfo {volume} {84}},\ \bibinfo {pages} {125303} (\bibinfo {year} {2011})}\BibitemShut {NoStop}%
\bibitem [{\citenamefont {Kotliar}\ and\ \citenamefont {Ruckenstein}(1986)}]{Kotliar1986}%
  \BibitemOpen
  \bibfield  {author} {\bibinfo {author} {\bibfnamefont {G.}~\bibnamefont {Kotliar}}\ and\ \bibinfo {author} {\bibfnamefont {A.~E.}\ \bibnamefont {Ruckenstein}},\ }\href@noop {} {\bibfield  {journal} {\bibinfo  {journal} {Phys. Rev. Lett.}\ }\textbf {\bibinfo {volume} {57}},\ \bibinfo {pages} {1362} (\bibinfo {year} {1986})}\BibitemShut {NoStop}%
\bibitem [{\citenamefont {Dong}\ and\ \citenamefont {Lei}(2001{\natexlab{a}})}]{BingDong2001}%
  \BibitemOpen
  \bibfield  {author} {\bibinfo {author} {\bibfnamefont {B.}~\bibnamefont {Dong}}\ and\ \bibinfo {author} {\bibfnamefont {X.~L.}\ \bibnamefont {Lei}},\ }\href@noop {} {\bibfield  {journal} {\bibinfo  {journal} {Phys. Rev. B}\ }\textbf {\bibinfo {volume} {63}},\ \bibinfo {pages} {235306} (\bibinfo {year} {2001}{\natexlab{a}})}\BibitemShut {NoStop}%
\bibitem [{\citenamefont {Dong}\ and\ \citenamefont {Lei}(2001{\natexlab{b}})}]{BingDong2001b}%
  \BibitemOpen
  \bibfield  {author} {\bibinfo {author} {\bibfnamefont {B.}~\bibnamefont {Dong}}\ and\ \bibinfo {author} {\bibfnamefont {X.~L.}\ \bibnamefont {Lei}},\ }\href@noop {} {\bibfield  {journal} {\bibinfo  {journal} {J. Phys.: Condens. Matter}\ }\textbf {\bibinfo {volume} {13}},\ \bibinfo {pages} {9245} (\bibinfo {year} {2001}{\natexlab{b}})}\BibitemShut {NoStop}%
\bibitem [{\citenamefont {Dong}\ and\ \citenamefont {Lei}(2002)}]{BingDong2002}%
  \BibitemOpen
  \bibfield  {author} {\bibinfo {author} {\bibfnamefont {B.}~\bibnamefont {Dong}}\ and\ \bibinfo {author} {\bibfnamefont {X.~L.}\ \bibnamefont {Lei}},\ }\href@noop {} {\bibfield  {journal} {\bibinfo  {journal} {Phys. Rev. B}\ }\textbf {\bibinfo {volume} {65}},\ \bibinfo {pages} {241304} (\bibinfo {year} {2002})}\BibitemShut {NoStop}%
\bibitem [{\citenamefont {Ding}\ and\ \citenamefont {Dong}(2003)}]{Ding2003}%
  \BibitemOpen
  \bibfield  {author} {\bibinfo {author} {\bibfnamefont {G.-H.}\ \bibnamefont {Ding}}\ and\ \bibinfo {author} {\bibfnamefont {B.}~\bibnamefont {Dong}},\ }\href {https://doi.org/10.1103/PhysRevB.67.195327} {\bibfield  {journal} {\bibinfo  {journal} {Phys. Rev. B}\ }\textbf {\bibinfo {volume} {67}},\ \bibinfo {pages} {195327} (\bibinfo {year} {2003})}\BibitemShut {NoStop}%
\bibitem [{\citenamefont {Vernek}\ \emph {et~al.}(2006)\citenamefont {Vernek}, \citenamefont {Sandler}, \citenamefont {Ulloa},\ and\ \citenamefont {Anda}}]{VERNEK2006608}%
  \BibitemOpen
  \bibfield  {author} {\bibinfo {author} {\bibfnamefont {E.}~\bibnamefont {Vernek}}, \bibinfo {author} {\bibfnamefont {N.}~\bibnamefont {Sandler}}, \bibinfo {author} {\bibfnamefont {S.}~\bibnamefont {Ulloa}},\ and\ \bibinfo {author} {\bibfnamefont {E.}~\bibnamefont {Anda}},\ }\href {https://doi.org/https://doi.org/10.1016/j.physe.2006.03.099} {\bibfield  {journal} {\bibinfo  {journal} {Physica E: Low-dimensional Systems and Nanostructures}\ }\textbf {\bibinfo {volume} {34}},\ \bibinfo {pages} {608} (\bibinfo {year} {2006})}\BibitemShut {NoStop}%
\bibitem [{\citenamefont {Ma}\ and\ \citenamefont {Lei}(2004)}]{JingMa_2004}%
  \BibitemOpen
  \bibfield  {author} {\bibinfo {author} {\bibfnamefont {J.}~\bibnamefont {Ma}}\ and\ \bibinfo {author} {\bibfnamefont {X.~L.}\ \bibnamefont {Lei}},\ }\href {https://doi.org/10.1209/epl/i2004-10079-7} {\bibfield  {journal} {\bibinfo  {journal} {Europhysics Letters}\ }\textbf {\bibinfo {volume} {67}},\ \bibinfo {pages} {432} (\bibinfo {year} {2004})}\BibitemShut {NoStop}%
\bibitem [{\citenamefont {Tsvelick}\ and\ \citenamefont {Wiegmann}(1983)}]{Tsvelick1983}%
  \BibitemOpen
  \bibfield  {author} {\bibinfo {author} {\bibfnamefont {A.~M.}\ \bibnamefont {Tsvelick}}\ and\ \bibinfo {author} {\bibfnamefont {P.~B.}\ \bibnamefont {Wiegmann}},\ }\href@noop {} {\bibfield  {journal} {\bibinfo  {journal} {Adv. Phys.}\ }\textbf {\bibinfo {volume} {32}},\ \bibinfo {pages} {453} (\bibinfo {year} {1983})}\BibitemShut {NoStop}%
\bibitem [{\citenamefont {Pruschke}\ and\ \citenamefont {Grewe}(1989)}]{Pruschke1989}%
  \BibitemOpen
  \bibfield  {author} {\bibinfo {author} {\bibfnamefont {T.}~\bibnamefont {Pruschke}}\ and\ \bibinfo {author} {\bibfnamefont {N.}~\bibnamefont {Grewe}},\ }\href@noop {} {\bibfield  {journal} {\bibinfo  {journal} {Z. Phys. B}\ }\textbf {\bibinfo {volume} {74}},\ \bibinfo {pages} {439} (\bibinfo {year} {1989})}\BibitemShut {NoStop}%
\bibitem [{\citenamefont {\v{Z}itko}(2021)}]{zitko_rok}%
  \BibitemOpen
  \bibfield  {author} {\bibinfo {author} {\bibfnamefont {R.}~\bibnamefont {\v{Z}itko}},\ }\href {https://doi.org/10.5281/zenodo.4841076} {\bibinfo {title} {{NRG Ljubljana}}} (\bibinfo {year} {2021})\BibitemShut {NoStop}%
\bibitem [{\citenamefont {Campo}\ and\ \citenamefont {Oliveira}(2005)}]{Campo2005}%
  \BibitemOpen
  \bibfield  {author} {\bibinfo {author} {\bibfnamefont {V.~L.}\ \bibnamefont {Campo}}\ and\ \bibinfo {author} {\bibfnamefont {L.~N.}\ \bibnamefont {Oliveira}},\ }\href {https://doi.org/10.1103/PhysRevB.72.104432} {\bibfield  {journal} {\bibinfo  {journal} {Phys. Rev. B}\ }\textbf {\bibinfo {volume} {72}},\ \bibinfo {pages} {104432} (\bibinfo {year} {2005})}\BibitemShut {NoStop}%
\bibitem [{\citenamefont {Hofstetter}(2000)}]{Hofstetter2000}%
  \BibitemOpen
  \bibfield  {author} {\bibinfo {author} {\bibfnamefont {W.}~\bibnamefont {Hofstetter}},\ }\href {https://doi.org/10.1103/PhysRevLett.85.1508} {\bibfield  {journal} {\bibinfo  {journal} {Phys. Rev. Lett.}\ }\textbf {\bibinfo {volume} {85}},\ \bibinfo {pages} {1508} (\bibinfo {year} {2000})}\BibitemShut {NoStop}%
\end{thebibliography}%

\end{document}